\documentclass[journal]{IEEEtran}
\usepackage{graphicx}
\usepackage{amsmath}
\usepackage{relsize}
\usepackage{amssymb}
\usepackage{subfigure}
\usepackage{multirow}
\usepackage{setspace} 
\usepackage{float}

\begin{document}

\title{2D Reconstruction of Small Intestine's Interior Wall}

\author{Rahman~Attar, Xiang~Xie, Zhihua~Wang, and Shigang~Yue

\thanks{R. Attar is with Centre for Computational Imaging \& Simulation Technologies in Biomedicine (CISTIB), Department of Electronic and Electrical Engineering, University of Sheffield, Sheffield, UK.}
\thanks{X. Xie and Z. Wang are with Institute of Microelectronics, Tsinghua University, Beijing, China.}
\thanks{S. Yue is with Computational Intelligence Laboratory (CIL), School of Computer Science, University of Lincoln, Lincoln, UK.}

\thanks{Manuscript received xxxx xx, xxxx;}}

\markboth{xxx}%
{Shell \MakeLowercase{\textit{et al.}}: Bare Demo of IEEEtran.cls for IEEE Journals}

\maketitle

\begin{abstract}
Examining and interpreting of a large number of wireless endoscopic images from the gastrointestinal tract is a tiresome task for physicians. A practical solution is to automatically construct a two dimensional representation of the gastrointestinal tract for easy inspection. However, little has been done on wireless endoscopic image stitching, let alone systematic investigation. The proposed new wireless endoscopic image stitching method consists of two main steps to improve the accuracy and efficiency of image registration. First, the keypoints are extracted by Principle Component Analysis and Scale Invariant Feature Transform (PCA-SIFT) algorithm and refined with Maximum Likelihood Estimation SAmple Consensus (MLESAC) outlier removal to find the most reliable keypoints. Second, the optimal transformation parameters obtained from first step are fed to the Normalised Mutual Information (NMI) algorithm as an initial solution. With modified Marquardt-Levenberg search strategy in a multiscale framework, the NMI can find the optimal transformation parameters in the shortest time. The proposed methodology has been tested on two different datasets - one with real wireless endoscopic images and another with images obtained from Micro-Ball (a new wireless cubic endoscopy system with six image sensors). The results have demonstrated the accuracy and robustness of the proposed methodology both visually and quantitatively - registration residual error of 0.93$\pm$0.33 pixels on 2500 real endoscopy image pairs and residual error accumulation of 16.59 pixels and without affecting the visual registration quality on stitching 152 images of Micro-Ball.
\end{abstract}


\IEEEpeerreviewmaketitle

\section{Introduction}

\IEEEPARstart{D}{evelopment} of wireless endoscopy has given gastroenterologists the opportunity to investigate the small intestine in a non-invasive way. Wireless capsule endoscopy is most commonly performed for obscure gastrointestinal (GI) bleeding, but also for other indications such as diagnosis of Crohn’s disease, suspicion of a small bowel tumour, diagnosis and surveillance of hereditary polyposis syndromes, non-steroidal anti-inflammatory drug-induced small intestinal lesions and celiac disease \cite{yuan2016bleeding, van2015}.
Wireless endoscopy offers several potential advantages over traditional esophagogastroduodenoscopy including the ability to perform 24 hours a day, without sedation, and interpreted at the bedside by physicians. In addition, it is painless and much less invasive, so enables the patient to pursue daily routines after the procedure. Therefore, it has a high patient acceptability rate \cite{meltzer2015}. However, interpretation procedure is time consuming and tiresome. Once swallowed, the device which typically has one or two cameras moves through the digestive tract and takes two or three pictures (depending on the model) per second for about eight hours. Thus, a large number of images are generated \cite{fu2014computer}. Moreover, low illumination and random movements of the device and therefore non-perfect shooting angles make the images less understandable. Besides, there are some symptoms (such as ulcer or polyp) that are usually spread over large areas in GI tract, whereas the endoscopic image area is very small \cite{weibel2010}. In order to detect these symptoms, it is crucial to build panoramic images of interior walls to extend the field of view \cite{bergen2016stitching}. Consequently, if a 2D representation map of the whole GI tract is possible to generate, it will be much easier to interpret rather than investigating the images individually.
However, it should be noted that the capsule has an average missing rate of $20\%\sim30\%$, due to limited visual field, random movements and low capture rate of the cameras, that imposes a challenge \cite{bergen2016stitching, meng2004}. To address the missing-rate issue, a smart wireless Micro-Ball endoscopy prototype system was proposed in \cite{gu2014}, where six cameras, constructing a cubic structure, are embedded in the Micro-Ball system. Determined by a smart control system, the device operates in monitor, normal and drastic modes with a capture rate of 1, 6 and 24 fps respectively. To reduce power consumption, a camera selection strategy is applied so that only one or two cameras work simultaneously without missing any areas.

\begin{figure}[t]
	\centering
	\includegraphics[width=2.7in]{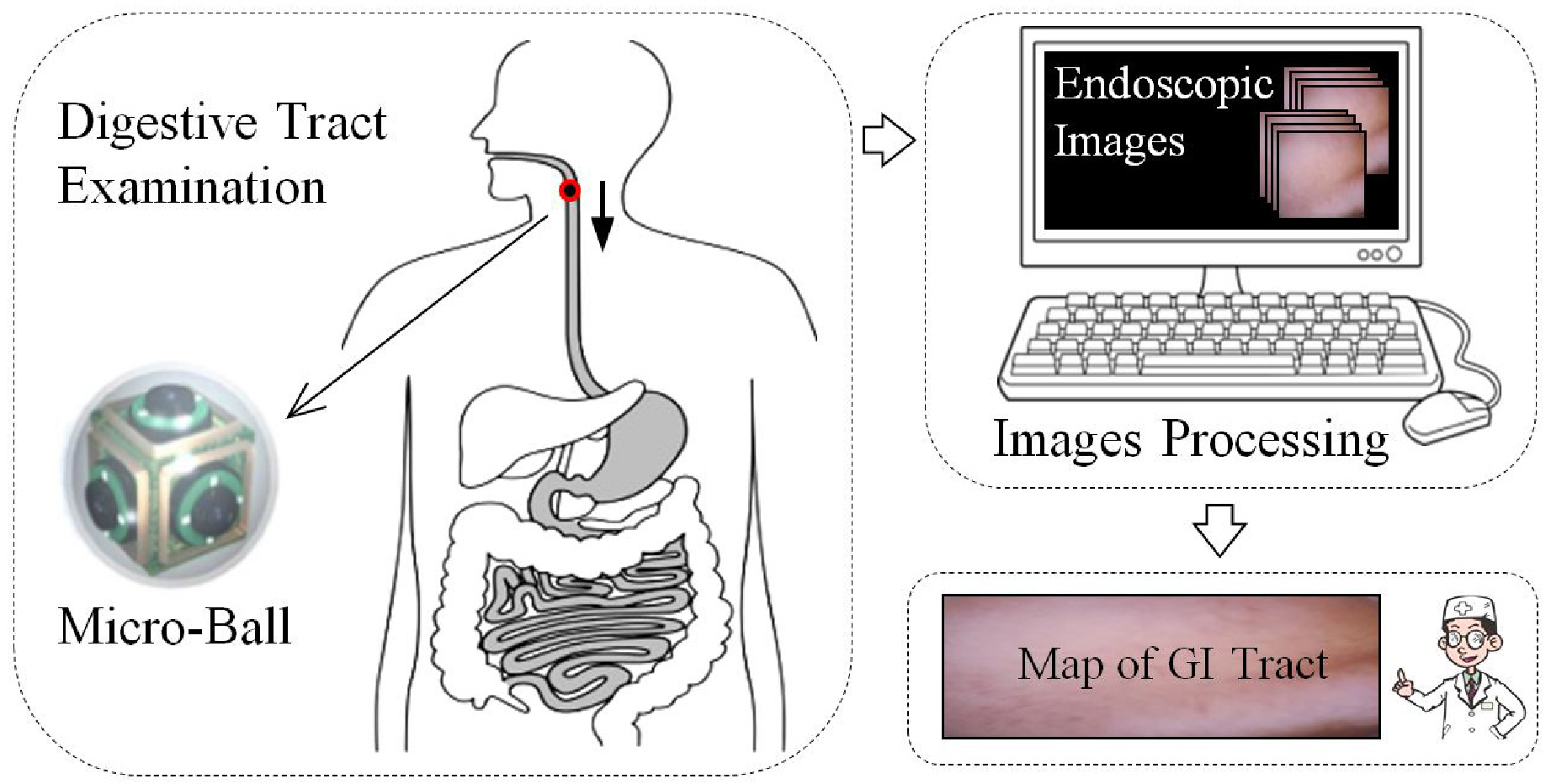}
	\caption{2D representation of GI tract images for clinical diagnosis.}
	\label{microball}
		\vspace{-0.6cm}
\end{figure}

As shown in Fig. \ref{microball}, the GI tract map can then be constructed successfully provided the images are efficiently registered. In this context, image registration is defined as follows: given two images, $I_A$ and $I_B$, to register the images is to find a optimal geometric transformation function $T_\mu$, where $T_\mu(I_A)$ best matches $I_B$, where $\mu$ is a set of transformation parameters.

Considering that the GI tract is soft and flexible together with the fact that the endoscopy device is propelled by gravity and peristaltic movements within the GI tract, reconstruction of the interior wall seems to be a formidable task and current algorithms fail to achieve satisfactory results within a reasonable time frame. Therefore, a new approach is required to overcome this challenge.

Recently attempts have been made to tackle the imposed challenges in \cite{wang2012} to create a 2D representation of GI tract with the Micro-Ball. This work has proposed a method of a straightforward combination of two image registration algorithms, i.e. Phase Correlation Method (PCM) and Scale Invariant Feature Transform (SIFT) \cite{lowe2004SIFT} to assess image registration algorithms on this particular application. However, in practice, further investigations are crucial to provide a better representation of the small intestinal tract for physicians.

The objective of the present work is to further improve the solution to the 2D reconstruction of the small intestine’s interior wall by proposing a new methodology to register images for easier inspection. Images are enhanced with a preprocessing method \cite{attar2014} to improve their quality. The keypoints are extracted with Principle Component Analysis and SIFT (PCA-SIFT) algorithm \cite{ke2004}, and then processed with Maximum Likelihood Estimation SAmple Consensus (MLESAC) \cite{torr1998} outlier removal to select the most correlated keypoints between paired images. The set of transformation parameters $\mu$ are considered as the candidate solutions of transformation function $T_\mu$. To improve the accuracy and efficiency of image registration, the optimal set of transformation parameters $\mu$ are tuned to maximize the Normalized Mutual Information (NMI) value between paired images. As becomes evident in the sequel, application of this technique on real wireless endoscopic images and also a unique database of images obtained from vitro experiments with the Micro-Ball demo system, shows the capability of reconstructing a 2D representation for visual inspection. The new proposed method works well for images of the current endoscopy devices, also is a perfect fit for the capsules with multiple cameras and larger field of view like the Micro-Ball.

The paper is organized as follows. In Section \ref{relatedWork}, endoscopic images and related research are reviewed briefly. The whole problem of 2D representation and the contribution of this paper is given in Section \ref{MethodologyContribution}. In Section \ref{proposed}, the image stitching algorithm is developed. The experimental results are presented in Section \ref{resultsDiscussion}. Finally, Section \ref{conclusion} concludes the paper and proposes future work.

\section{Endoscopic Image Analysis and Related Work}
\label{relatedWork}
\begin{figure}[t]
	\centering
	\includegraphics[width=3.5in]{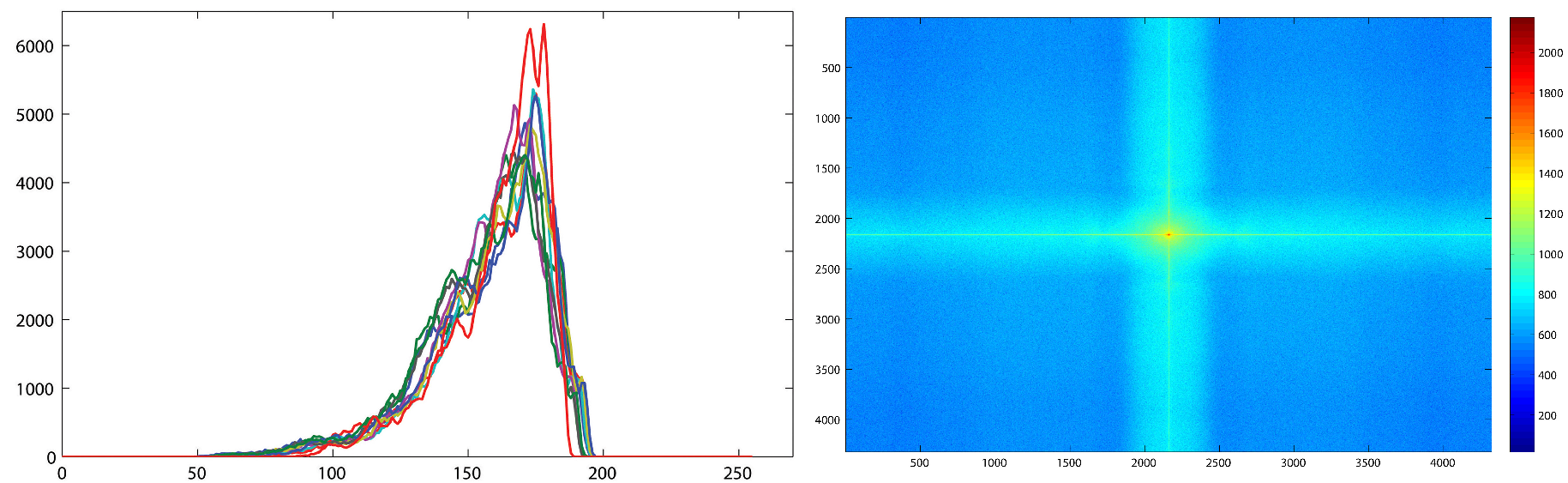}
	\caption{Histogram (left) and Frequency analysis (right) of 10 sample endoscopic images.}
	\label{hist}
		\vspace{-0.6cm}
\end{figure}

Bad imaging conditions such as low illumination, high compression ratio, limitations in power consumption, short-focal-length camera, communication bandwidth, and storage device all affect the image quality \cite{yuan2016bleeding, bergen2016stitching, barbalata2016laryngeal}.
Observing a large number of histograms of endoscopic images, it can be seen that distribution of pixels concentrates in a certain range of intensity bins. Fig. \ref{hist} (left) shows distribution of pixels for ten samples. Furthermore, the most striking result to emerge from investigating all pixels of whole images in the dataset (including 89,800 images from two subjects) is that 81.86\% of pixels fall in 28.12\% of intensity bins. Moreover, the frequency domain analysis using Discrete Fourier Transform (DFT) for ten typical images of the GI tract is shown in Fig. \ref{hist} (right). In the frequency domain image, each point represents a particular frequency contained in the spatial domain image. The number of high frequency components are too small. It means that the features are not distinct and it brings challenges both for medics to interpret and engineers to process them for different applications.

As noted earlier, thousands of images with the aforementioned features are recorded for physicians to inspect. Depending on the experience of the examiner, the average time required for the visual inspection ranges from 45 to 120 minutes. This will result in an unnecessary waste of time and efforts. In literature, apart from those methods that directly detect abnormal frames (such as \cite{yuan2016bleeding, barbalata2016laryngeal, zhang2017automatic, segui2014detection}), there are two main groups of algorithms that have been proposed to reduce the visual inspection time, yet improve the quality of examination.

First group reduce the number of images by removing redundancy from dataset. For example: In \cite{karargyris2009}, a registration methodology has been proposed to reduce video length by removing same or similar frames. It utilizes a segmentation scheme and a graph method to register similar regions in two consecutive frames. In a similar work \cite{weibel2010}, it is shown that for image registration, the sparse graph cut method yields a computation speed up by an order of magnitude in comparison to a standard dense graph cut. In a different study \cite{lee2013}, a combination of an intensity correction method with a method based an optical flow and features is proposed to detect and reduce near-duplicate images. With the same objective in \cite{iakovidis2010}, an approach is proposed based on a data reduction algorithm to extract representative video frames from the full length endoscopic images. 

Second group use image processing and computer vision algorithms to produce clearer, more detailed, and wider view of region of interest. That is to say, information from a set of images introduce a 2D/3D informative representative. For example: In \cite{miranda2008}, a method has been proposed to facilitate clinical diagnosis by building a panoramic image of the bladder using images acquired from different viewpoints. Consecutive images are pairwise registered and then all are placed in a common and global coordinate system. More recently, a method for automatic construction of panoramic visual summaries was proposed in \cite{spyrou2013}. It involves extraction of visual features from each frame, and frame registration by Speeded Up Robust Features (SURF) algorithm.  According to an investigation by \cite{karargyris2011}, methodologies are proposed that can create new interpolated frames for smoothing video representation and having a more realistic 3D view of a few regions of the digestive tract surface, but not a 3D representation of the whole tract. In \cite{szczypinski2009}, a method is put forward based on the model of deformable rings. The model processes the video to characterize capsule's motion and to produce a map - a rough representation of the GI system's endoluminal surface. The novelty of the model lies in the shape of its mesh, designed particularly to follow the specific type of motions. 

However, in practice, further investigations are needed to provide a better solution. This work focuses on developing an algorithm to address concerns of both above groups, i.e. reducing the number of images in the dataset as well as providing a wider representation of region of interest for easier inspection.
	\vspace{-0.2cm}

\section{Methodology and Contribution}
\label{MethodologyContribution}
Generally, image registration algorithms can be classified into feature-based and intensity-based methods. 

Feature-based methods establish a correspondence between a number of especially distinct points (using keypoints detector algorithms) in images, then a geometrical transformation is determined to map the target image to the reference images. One of the main advantages of these approaches is that they are fast and robust to noises, complex geometric distortions, and significant radiometric differences. Supremacy of SIFT in comparison with other keypoint detectors/descriptors is reported in \cite{bauer2007, mikolajczyk2005}. Although SIFT performs best for most of the cases it is unable to provide satisfactory result when it's directly applied to the endoscopic images. Thus, using SIFT alone cannot produce optimal results because there are only a small number of correct matches, keypoints are not evenly spread over the image, and many outliers exist in feature matches. 

Intensity-based methods, which are more popular in medical image registration, employ the intensity of pixels. Mutual Information (MI) is the most popular measure because of its robustness to intensity variation and noise, and high accuracy. It indicates any linear/non-linear correlation \cite{shams2010}, and produces consistently
sharper peaks at the correct registration values than other intensity-based methods,
which is very important for accurate registration. Despite the general promising results, using MI can result in misregistration when the images have low resolution or when the images contain little information \cite{pluim2000, thevenaz1997}. However, the registration algorithm performs much better when it starts maximisation of the Normalised MI (NMI) from an accurate initial solution which is close to ground truth.

Therefore, in this work, a hybrid algorithm is put forward based on the combination of an intensity-based and a feature-based method that has shown the best performance in images registration. The advantages of both methods are employed to complement each other. Based on the experiments that have been done on endoscopic images in this work, Normalized Cross Correlation (NCC) which is an intensity-based method produces very similar results to NMI. When NMI is replaced with NCC in the algorithm the overall error is slightly more (1.27\%). SURF (a feature-based method) also was tested instead of SIFT. Although SURF is faster than SIFT (37.88\%) the results obtained by using SIFT is more accurate (14.71\%). All these results declare that above combination of the methods works the best on endoscopic images but someone may get better results in future works by changing/modifying the method in the initialization step and/or fine-tuning step. The structure of the algorithm presented here is in a way that initialization step can be done with a feature-based method and then an intensity-based method is used to tune the transformation function.

\begin{figure}[!t]
	\centering
	\includegraphics[width=3in]{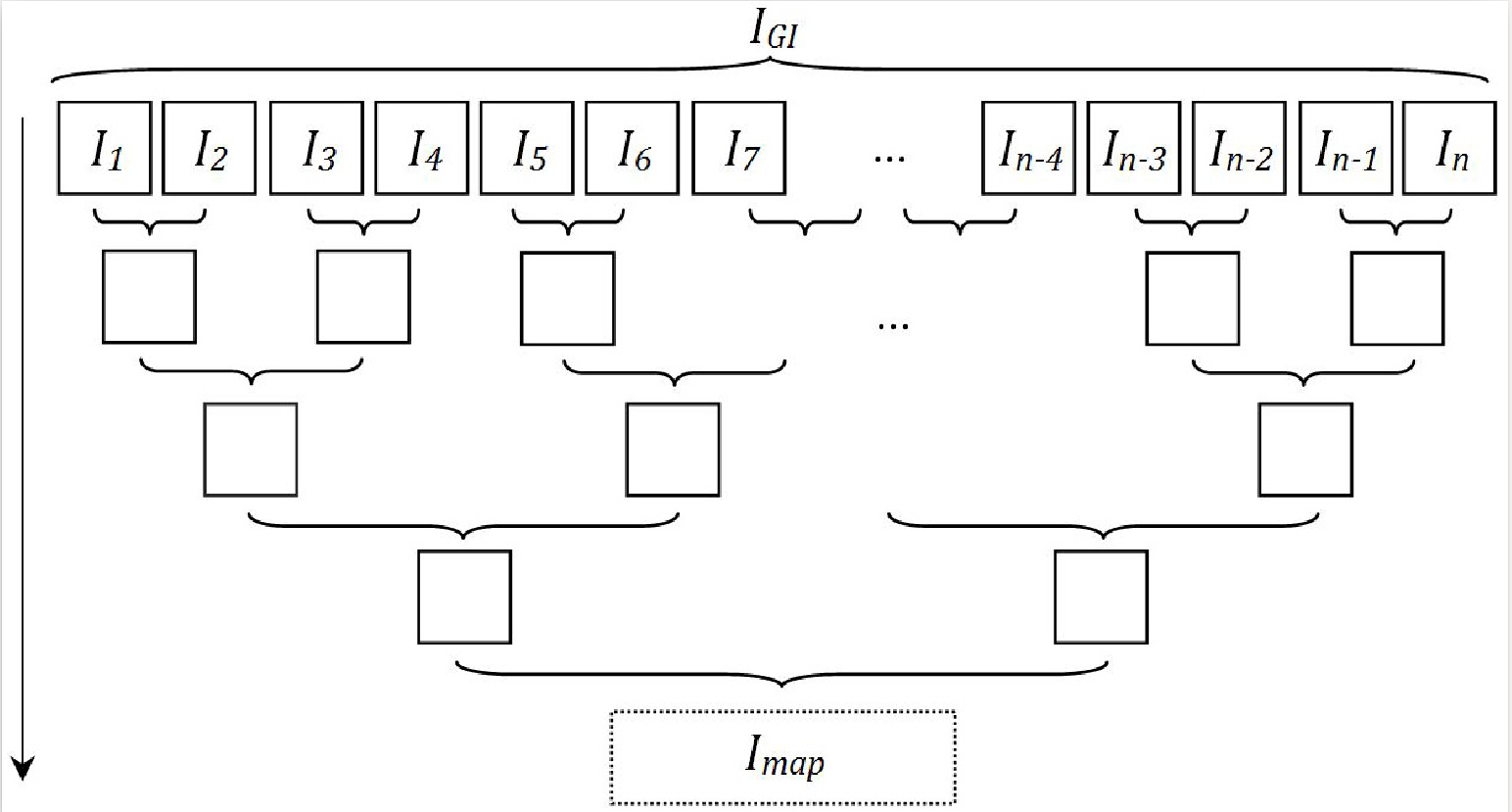}
	\caption{Scheme of endoscopic image registration of each two adjacent images.}
	\label{registration}
		\vspace{-0.6cm}
\end{figure}

The problem of registration is being solved for any two adjacent images of input image dataset: $I_{GI}~=~\{I_1, I_2, . . . ,I_n\}$,  it gradually develops along the dataset to form the GI tract map ($I_{map}$), as illustrated in Fig. \ref{registration}.

Every image registration algorithm requires three basic ingredients: a spatial transformation model which determines the set of possible solutions, an objective similarity measure which estimates the quality of each potential solution, and an optimization algorithm which looks for the best solution \cite{maintz1998}. Therefore, the whole procedure is to solve the image registration problem formulated by the following equations for any two adjacent images.
\begin{equation}‎
\tilde T_\mu=‎arg~opt_{T_\mu}~sim\big((\Omega_{A})~,~T_\mu(\Omega_{B})\big)
\label{stage2}
\end{equation}‎\begin{equation}‎
\Omega_{A}=f_{A}\big(I_A^{ov}(d_{GI} ,d_{CA})\big)
\label{stage2.2}
\end{equation}\begin{equation}‎
\Omega_{B}=f_{B}\big(I_B^{ov}(d_{GI}^{~'},d_{CA}^{~'})\big)
\label{stage2.3}
\end{equation}
where $\mu$ is a set of transformation parameters which can lead to rotation, scale, shear, shift, and projection between the two images. Due to the small surface deformation, highly overlapped frames, and slow movement of the camera there are only small displacements between consecutive images of an endoscopic video sequence. Therefore it is possible to successfully register image pairs using 2D perspective transformations \cite{miranda2008, weibel2010}. $I_{A}^{ov}$ and $I_{B}^{ov}$ are the overlapping parts of two successive images $I_{A}$ and $I_{B}$, respectively. Usually, in image registration, homologous information $\Omega_{A}$ and $\Omega_{B}$ are extracted from images using a particular algorithm $f$. However, due to lack of enough distinct features in endoscopic images, it's difficult to select $\Omega$ and $f$. Meanwhile, image deformation $d_{GI}$ (caused by soft tissue deformation of GI tract) and $d_{CA}$ (caused by random motion of the endoscopy device) have negatively affected the image registration process. Generally, the homologous structures are superimposed by an optimization procedure $opt_{T_\mu}$, which maximizes the similarity function $sim$, or minimise dissimilarity.  

\begin{figure}[!t]
	\centering
	\includegraphics[width=2.8in]{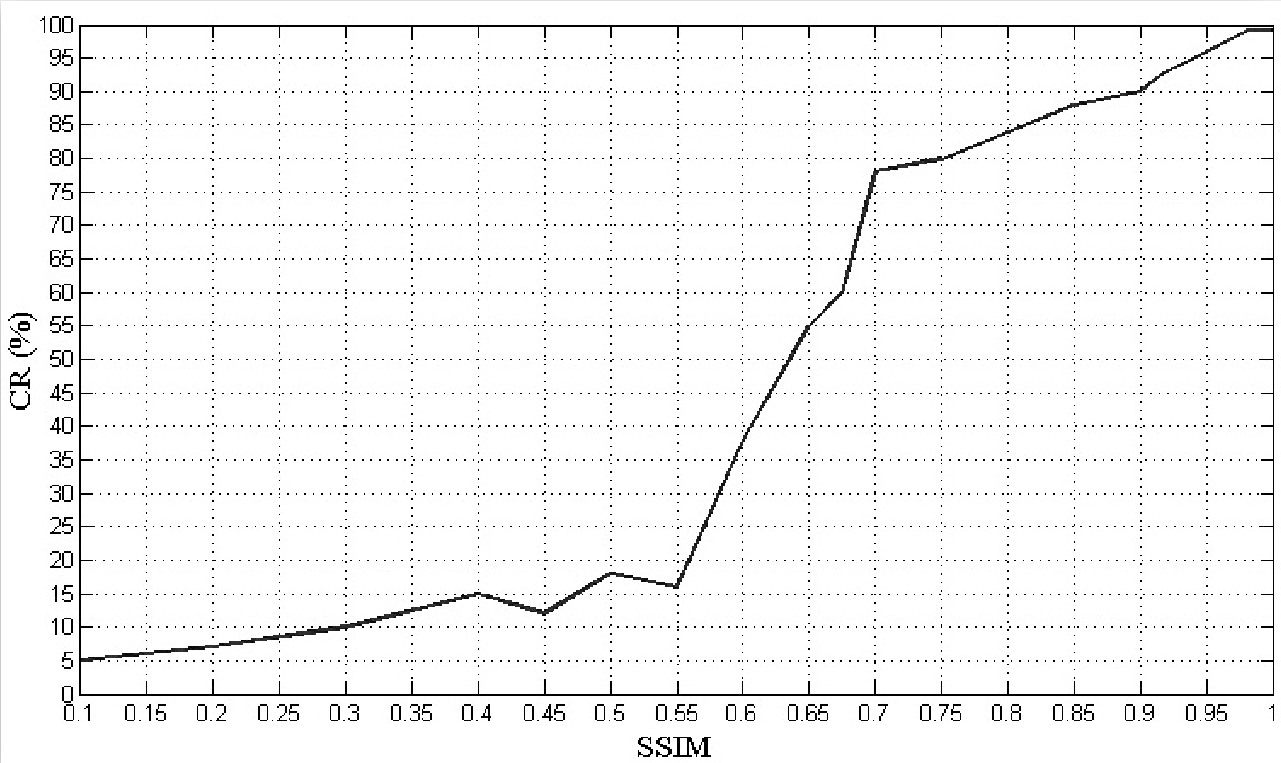}
	\caption{Behaviour of SSIM vs CR; higher SSIM shows higher correct rate.}
	\label{ssimD}
	\vspace{-0.6cm}
\end{figure}

Due to the unpredictable motions (i.e. bouncing) of the capsule camera, the probability that two successive frames are very dissimilar is relatively high. In such cases it is crucial that the methodology should not perform registration. In order to handle these cases, to create a decision-making model, a simple check between two successive images by Structural SIMilarity ($SSIM$) index is used. $SSIM$ basically shows the degree of similarity of two images, scaled on the range $0 \leq SSIM \leq 1$, the higher the value of $SSIM$, the more similarity there is \cite{attar2015, wang2004}. Therefore, the registration method is applied to those image pairs whose $SSIM$ is greater than a threshold ($th$). According to the Fig. \ref{ssimD} which shows Correct Rate (CR; explained in Section \ref{resultsDiscussion}) versus $SSIM$, the threshold is set to $th=0.70$. Clearly, for the values lower than $th$, the CR is too low and for the values greater than $th$, CR increases. Thus, based on this predictable behaviour a judgement is possible.

\begin{figure}[!t]
	\centering
	\includegraphics[width=3in]{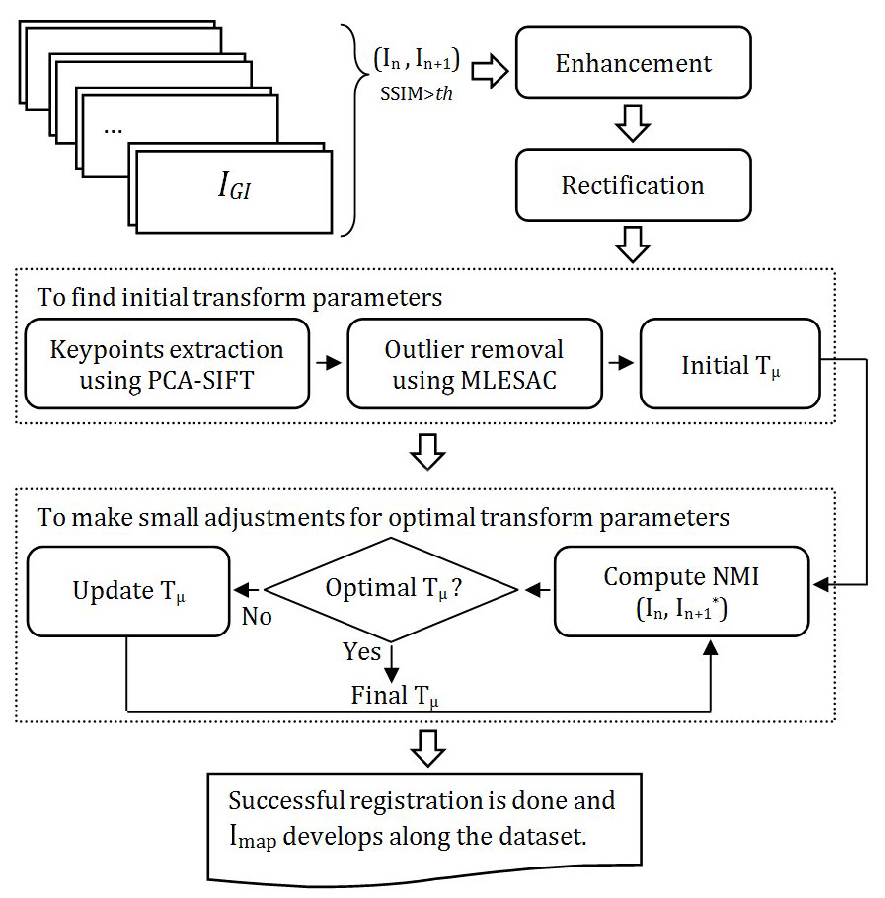}
	\caption{Flowchart of the proposed method. $I_n$ and $I_{n+1}$ = two consecutive images; $I_{n+1}^*$ = transformed image; $T_{\mu}$ = transform function.}
	\label{diagram}
		\vspace{-0.6cm}
\end{figure}

Consequently, the hybrid registration scheme proposed in this work, after preprocessing, is composed of the following main stages, visualized in Fig. \ref{diagram}. To begin with, an initial solution for the transform function is obtained through a preregistration procedure based on scale invariant keypoints which have been extracted and refined with a reliable outlier removal. Then, fine tuning of the registration parameters is accomplished in order to maximize NMI over each paired images. The main innovation and contribution of this study are summarised as follows:

1) This work proposes an efficient automatic algorithm that employs the robustness of feature-based methods and the accuracy of intensity-based methods to construct a hybrid scheme for endoscopic image registration.

2) By examining the performance of the best feature-based keypoints detectors/descriptors algorithm \cite{lowe2004SIFT}, i.e. SIFT algorithm and its two variants (PCA-SIFT and SIFT+GC) with the most widely used outlier removal algorithms (i.e. HT, RANSAC, MSAC and MLESAC) on wireless endoscopic images dataset, the best couple is selected to produce the best primary results to guarantee that the preregistration results are close enough to the ground truth.

3) For maximization of NMI, an excellent initial solution selection strategy is put forward by using PCA-SIFT with MLESAC outlier removal at the coarsest level of search stage. Registration accuracy has improved significantly compared with standard NMI which uses a random initial solution.

4) A modification of Marquardt-Levenberg search strategy in a multiscale framework is exploited to refine the coarser level registration results and reach the most precise registration parameters for maximising NMI.

5) The proposed registration strategy, besides common wireless endoscopic images, is tested on a unique database which has been obtained from \textit{in vitro} experiments with the Micro-Ball demo system. 

In the following sections, different steps are separately described and discussed in detail. 

\section{Proposed Method}
\label{proposed}
The proposed algorithm, as illustrated in Fig. \ref{diagram}, consists of the following steps: 
	\vspace{-0.3cm}
\subsection{Image Enhancement}
As an efficient preprocessing step to enhance the quality of endoscopic images to make their feature ($\Omega$) more distinct and easier to examine, a simple approach proposed in \cite{attar2014} is used. This method by histogram information correction and edge preserving mask accentuate features of the interest in endoscopic images. 
	\vspace{-0.3cm}
\subsection{Image Rectification}
Rectification is the process of transforming an acquired image into a rectified image, for which the resulting overall transformation from physical object to rectified image has the ideal form. The distortion correction problem can be solved by an efficient approach as a separate preprocessing step to improve the quality of registration.

To overcome the issue of $d_{CA}$ which is caused by random motion of the Micro-Ball, perspective distortion of wireless endoscopic images needs to be rectified. There are several rectification methods using the advantage of reference corners in images, but it is not easy to automatically find such keypoints in wireless endoscopic images. Fortunately, the location information given by the attitude-sensing system embedded in the Micro-Ball can be used to solve the image rectification problem \cite{hu2011attitude}. In addition to that sensor which includes a magnetometer and an accelerator, there is another attitude-sensing sensor which is worn on the patient body and it consists of magnetometer, accelerator, and gyroscope. Using these two attitude sensing systems, the intestine's orientation can be driven, more details are available in \cite{gu2014, hu2011attitude}.

\begin{figure}[!t]
	\centering
	\includegraphics[width=2.7in]{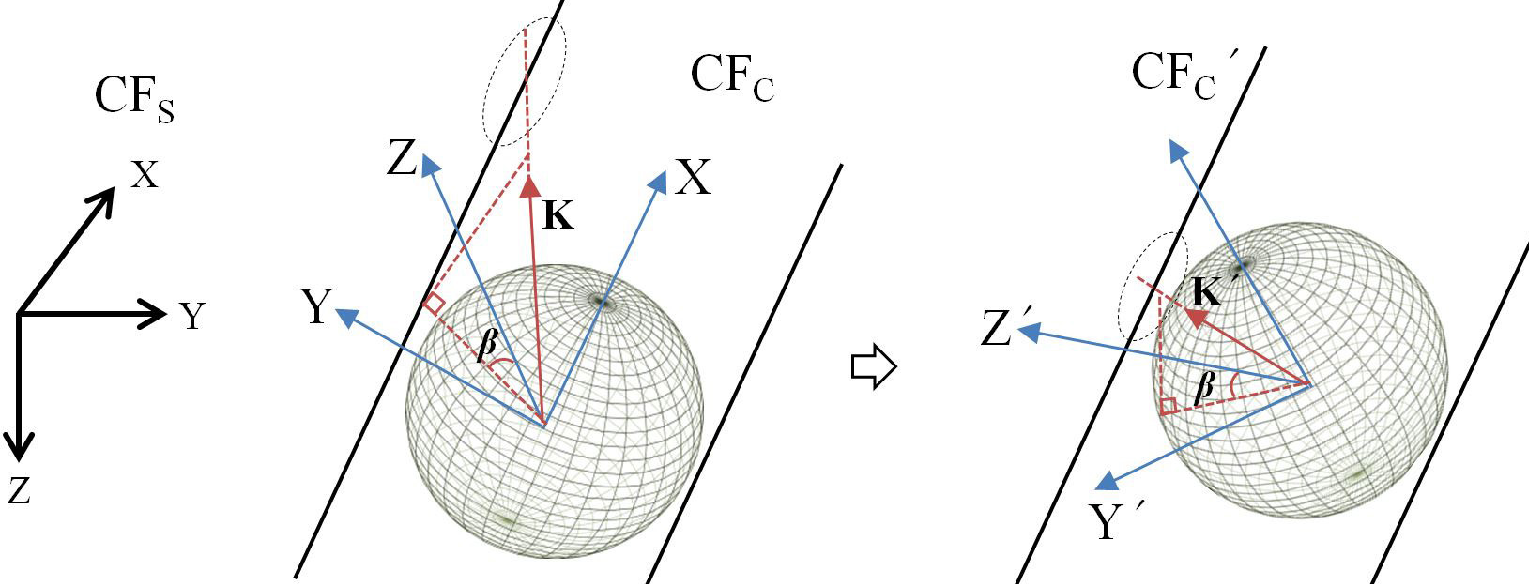}
	\caption{The diagrammatic sketch of perspective distortion correction.}
	\label{rect1}
		\vspace{-0.6cm}
\end{figure}

The distortion correction process is briefly shown in Fig. \ref{rect1}. We do image correction to find actual relationship $V(\theta,\varphi,\alpha,focal)$ between pre-corrected image ($CF_c$) and post-corrected image ($CF_c'$). $\theta$, $\varphi$, $\alpha$, and $focal$ are, respectively, roll, pitch, yaw, and the focal length of the camera, available from Micro-Ball device. $CF_s$ is the coordinate system relative to the practical plane. Because of the invariability of any physical point coordinates in the real world, then $V(\theta,\varphi,\alpha,focal)$ can be obtained.

\begin{figure}[!t]
	\centering
	\includegraphics[width=2in]{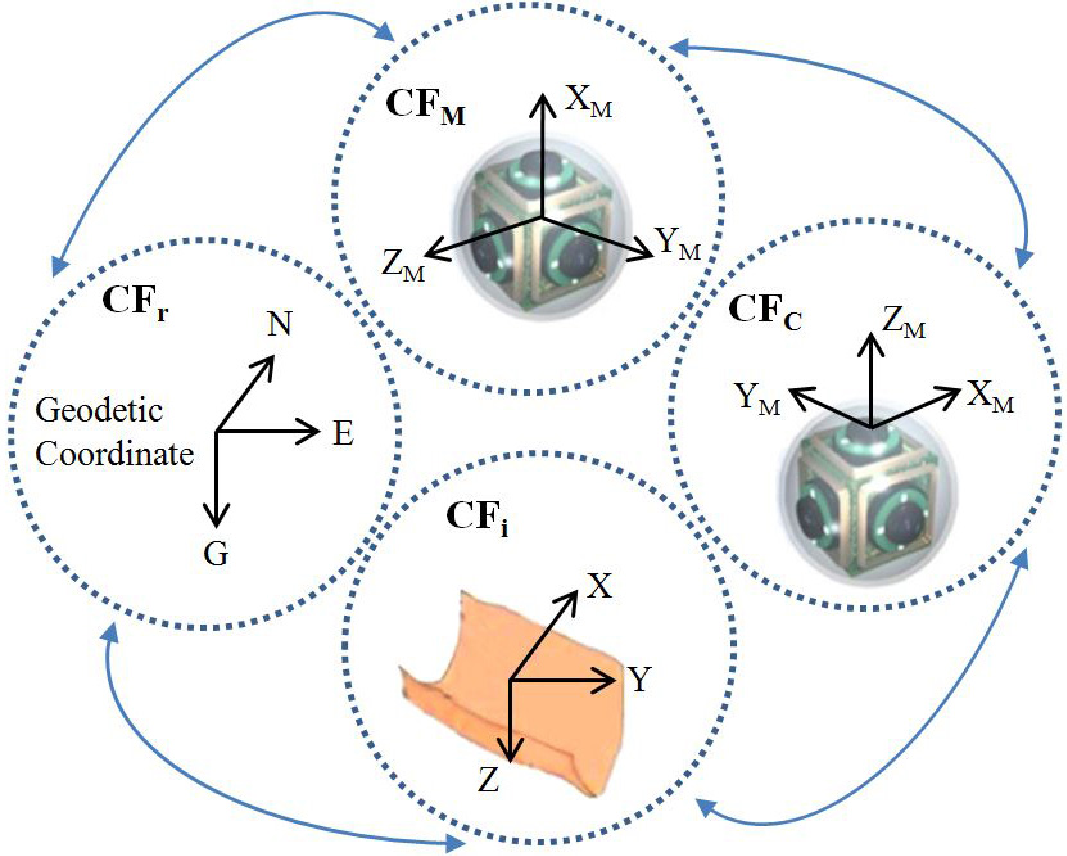}
	\caption{Coordinate systems in image correction for Micro-Ball images.}
	\label{rect2}
		\vspace{-0.5cm}
\end{figure}

In the Micro-Ball endoscopy system, as the captured image plane is changeable and each camera has its own coordinate system, the problem of image geometry correction is more complex. As shown in Fig. \ref{rect2}, the coordinate systems are illustrated as follows:
\\$CF_r$: reference frame of attitude-sensing system;
\\$CF_M$: carrier coordinate system of the Micro-Ball;
\\$CF_c$: frame of the camera to be corrected;
\\$CF_i$: frame of the intestine;
\\$CF_s$: frame corresponding to the image plane captured.

As mentioned earlier, the relationship between $CF_c$ and $CF_s$ must be acquired to obtain image correction, and it is possible through the following four steps:

Step 1: $CF_M=R.CF_r$ ; $R$ is the rotation transform matrix given by the endoscopy device.

Step 2: $CF_c=T_1.CF_M$ ; This step is essential, because $CF_M$ is fixed but $CF_c$ is changing with the camera to be corrected.

Step 3: $CF_r=T_2.CF_i$ ; This relationship is acquired based on the intestinal lumen detection combined with coordinate geometry location information from the Micro-Ball.

Step 4: $CF_i=T_3.CF_s$. As shown in Fig. \ref{rect1}, the shooting direction of the camera, vector $K$ in $CF_i$, is denoted as $k_{nt}$ 
\begin{equation}‎
k_{nt}=T^{-1}_{2}R^{-1}.k_{n0}\quad n=1, 2, . . . , 6
\label{knt}
\end{equation}‎where $k_{n0}$ is $K$'s vector representation in $CF_M$ are as follow\\\\
$ \begin{array}{lcr}
k_{10}=(0,0,-1)^T&k_{20}=(0,0,1)^T&k_{30}=(1,0,0)^T\\
k_{40}=(0,-1,0)^T&k_{50}=(-1,0,0)^T&k_{60}=(0,1,0)^T
\end{array} $
\\\\
\noindent ‎With $k_{nt}$, the angle $\beta$ in Fig. \ref{rect1} can be calculated and then $T_3$ can be obtained by 
	\vspace{-0.1cm}
{\small \begin{equation}‎
	T_3=
	\begin{bmatrix}
	1&0&0\\
	0&cos\beta&-sin\beta\\
	0&sin\beta&cos\beta
	\end{bmatrix}^T
	\label{T3}
	\end{equation}‎}

\begin{figure}[!t]
	\centering
	\includegraphics[width=2.5in]{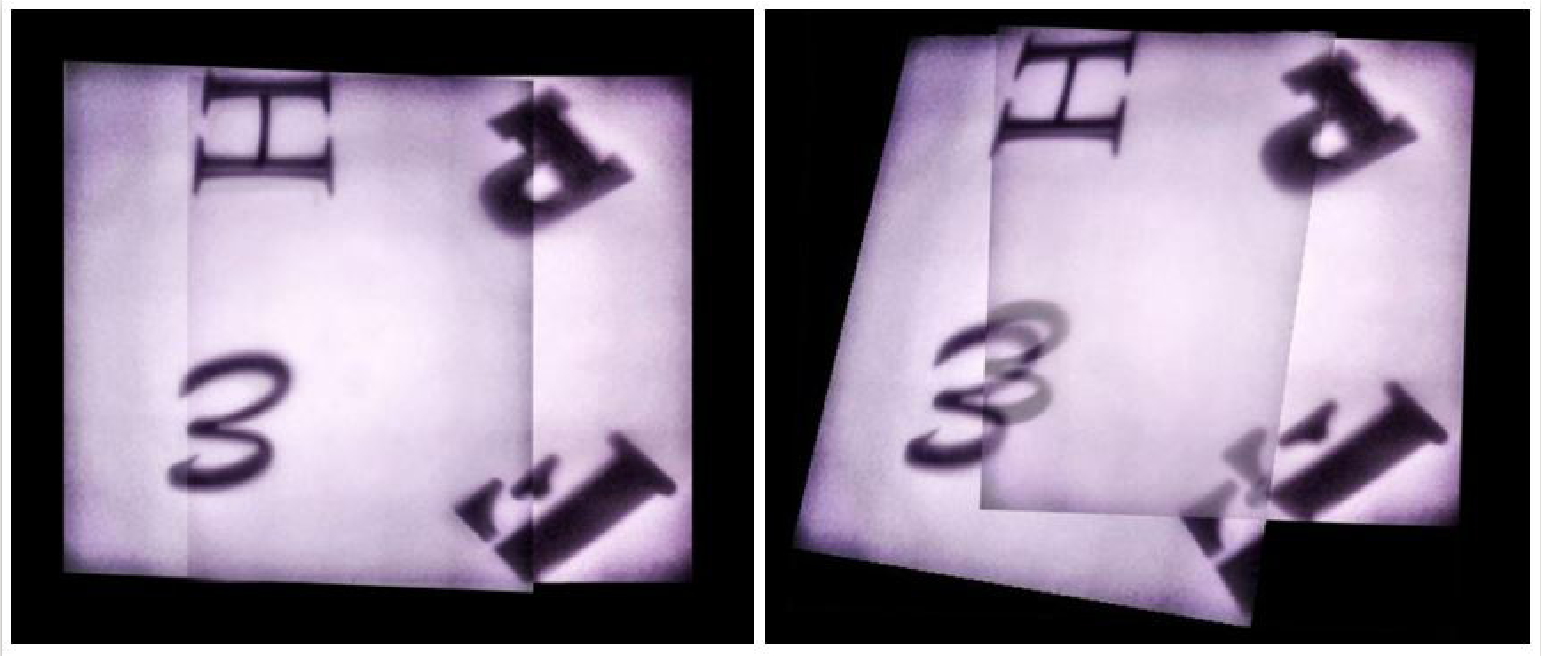}
	\caption{Result of registration with and without rectification method. Images are captured through simulation experiment of Micro-Ball which is explained in section \ref{resultsDiscussion}.}
	\label{with}
		\vspace{-0.5cm}
\end{figure}

According to above-mentioned steps, the relationship between $CF_c$ and $CF_s$ can be obtained by
	\vspace{-0.1cm}
\begin{equation}‎
CF_c=T_1.R.T_2.T_3.CF_s
\label{CF}
\end{equation}‎
	\vspace{-0.1cm}
Moreover, $T_1.R.T_2.T_3$ can be used to obtain $V(\theta,\varphi,\alpha,focal)$ and realize image correction. Although the image perspective distortion correction is made before the registration process the following perspective transformation function is used to have more accurate results.
	\vspace{-0.1cm}
\begin{equation}‎
T_\mu(x,y)=
\begin{bmatrix}
a_{11} & a_{12} & a_{13}\\
a_{21} & a_{22} & a_{23}\\
a_{31} & a_{32} & 1
\end{bmatrix}
\begin{bmatrix}
x\\
y\\
1
\end{bmatrix}
\label{transform}
\end{equation}‎ 
	\vspace{-0.1cm}
Fig. \ref{with} shows the result of registration of an image pair with and without image perspective distortion correction. Thus, perspective distortion correction for registration of images obtained through Micro-Ball is a crucial step.

	\vspace{-0.3cm}

\subsection{Scale-Invariant Feature Transform}
This section provides a brief overview of SIFT algorithm and its two variants namely PCA-SIFT \cite{ke2004} and SIFT+GC \cite{mortensen2005}. Following are different steps of SIFT algorithm for generating keypoints and their descriptors: 1) Scale Space creation
2) Laplacian of Gaussian (LoG) Approximation
3) Keypoints detection
4) Keypoints elimination
5) Keypoints orientation
6) Descriptors formation.

After the keypoint descriptor has been calculated, keypoints are matched by using the minimum distance method. However, not every pair of matched keypoints will be useful. 

In this work, two different editions of SIFT also are examined to find the best result for the application, in terms of time efficiency and accuracy; SIFT+GC and PCA-SIFT; SIFT+GC integrates global information into SIFT and PCA-SIFT applies PCA to the normalized gradient patch. In Section \ref{resultsDiscussion}, the performance of three above scale invariant keypoints detectors are investigated and PCA-SIFT is selected for the application as it has better performance than standard SIFT or SIFT+GC.
	\vspace{-0.3cm}
\subsection{Mismatch Detection}
\begin{figure}[!t]
	\centering
	\includegraphics[width=3.4in]{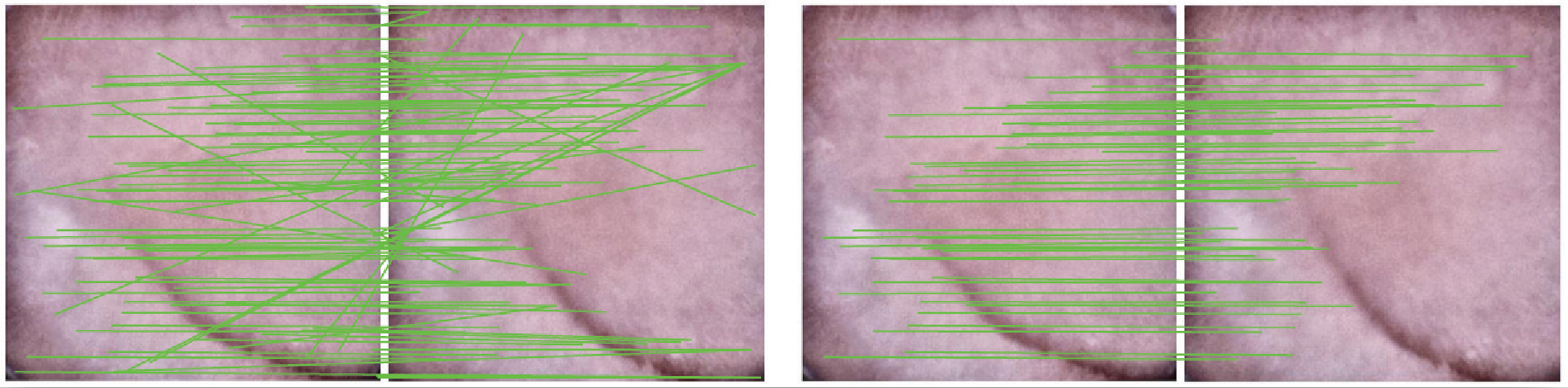}
	\caption{Matches found without and with applying external outlier removal.}
	\label{mismatches}
		\vspace{-0.5cm}
\end{figure}

Although PCA-SIFT has a very good performance, when it is used alone for image registration, many false matches arise. These false matches lead to further incorrect geometric correction. Therefore, a reliable outlier removal procedure is needed to ensure correct registration.
As an example, Fig. \ref{mismatches} shows corresponding keypoints between an image pair which are extracted and matched by PCA-SIFT algorithm and then mismatches are removed by an outlier removal.

Even small estimation errors can reduce the efficiency of the selection method drastically. In practice, the optimal set of features is usually unknown, and it is common to have irrelevant or redundant features at the beginning of process. 

Sum of squared differences between corresponding keypoints is used to measure the strength of candidate matches. Because the matching process is only based on proximity and similarity, mismatches often occur. The threshold for match acceptance is deliberately conservative at this stage to minimise incorrect matches. Consequently robust methods must also be adopted, which can provide a good estimate of the solution even if some of the data are mismatches (outliers). Correct matches will obey the epipolar geometry. The aim then is to obtain a set of inliers consistent with the epipolar geometry using a robust technique. In this case outliers are putative matches inconsistent with the epipolar geometry.

The first criterion is $T_{th}={d_1}/{d_2}$, an indicator of the ambiguity of each matching; where $d_1$ is the Euclidean distance in feature space between a keypoint in the input image and its nearest matched keypoint in the other image and $d_2$ is the Euclidean distance to the second nearest keypoint. If $T_{th}$ is close to one, it means $d_1$ is close to $d_2$. That is to say, for a certain keypoint in the input image, algorithm has detected two possible matching keypoints in the reference image. This is an ambiguous situation, and the matched pair will be deleted if $T_{th} > 0.80$. The second criterion is a spatial constraint based on an external algorithm. Following algorithms are examined one by one as second step to eliminate falsely matched keypoints using the spatial relationship: HT \cite{hough1962}, RANSAC \cite{fischler1981}, MSAC \cite{torr2000}, and MLESAC \cite{torr1998}.

Consequently, performance of three different scale invariant keypoints detectors (SIFT, PCA-SIFT, and SIFT+GC) accompanied with four different outlier removal algorithms (HT, RANSAC, MSAC, and MLESAC) are examined one by one, i.e. totally 12 different cases. The results show that a combination of PCA-SIFT and MLESAC has the best performance among others for the application. In Section \ref{resultsDiscussion}, the experimental results are shown. 

Based on the experimental results, the most reliable keypoints that have been extracted by PCA-SIFT-MLESAC are used to form the initial transform function $T_\mu$ so as to be used in the next step of the proposed method for further improvement.

	\vspace{-0.3cm}
\subsection{Mutual Information}
There are two main characteristics that distinguish MI from other dependency measures: 
\textit{First}, the capacity of measuring any kind of relationship between variables;
\textit{Second}, its invariance under space transformations. MI between two images takes its maximum value when images are absolutely dependent. Registration is assumed to correspond to maximizing MI: the images have to be aligned in such a manner that the amount of information they contain about each other is maximal.

\subsubsection{MI Formulation}
Giving two images to register, $f_A(x,y)$ and $f_B(x,y)$, defined on a continuous domain $V^c$, the coordinates $(x_i,y_i)$ are samples of $V^c$, and the discrete set of these samples is called $V$. Let $L_A$ and $L_B$ be discrete sets of intensities of two images, and $T(\mu)$ be the geometric transformation with associated parameters $\mu$. $MI$ is 
	\vspace{-0.1cm}
\begin{equation}‎
MI(\mu)=\sum_{a\in L_A}\sum_{b\in L_B}p(a,b;\mu)~.~log_2(\frac{p(a,b;\mu)}{p_A(a;\mu)~p_B(b;\mu)})
\label{mutual}
\end{equation}

\noindent ‎where $p_A(a;\mu)$ and $p_B(b;\mu)$ are the marginal probability distribution, $p(a,b;\mu)$ is the joint probability distribution, $a\in L_A$, and $b\in L_B$. 

While $MI$ measures the degree of dependency of $f_A(x,y)$ and $f_B(x,y)$ by measuring the distance between the joint distribution $p(a,b;\mu)$ and the distribution associated to the case of complete independence $p_A(a;\mu)~.~p_B(b;\mu)$ through Kullback–Leibler measure \cite{vajda1989theory}. These distributions can be obtained by
	\vspace{-0.1cm}
\begin{equation}‎
p(a,b;\mu)=\frac{h(a,b;\mu)}{\displaystyle \sum_{a\in L_A}\displaystyle \sum_{b\in L_B}h(a,b;\mu)}
\label{distribution1}
\end{equation}‎ 
	\vspace{-0.1cm}
\begin{equation}‎
p(a;\mu)=\sum_{a\in L_A}p(a,b;\mu)~ and~ p(b;\mu)=\sum_{b\in L_B}p(a,b;\mu)
\label{distribution2}
	\vspace{-0.1cm}
\end{equation}

\noindent ‎where $h$ is the joint discrete Parzen histogram given by (\ref{parzen}), the Parzen histogram is built based on the partial overlapping domain of $f_A(x, y)$ and $f_B(T_\mu(x,y))$, namely, the gray values $(a,b)$ of those pairs of pixels which lie in the same position.
	\vspace{-0.1cm}
\begin{equation}‎
\begin{array}{l l}
h(a,b;\mu)=\displaystyle \frac{1}{\varepsilon_A \varepsilon_B}\displaystyle \sum_{(x_i,y_i)\in V} \omega(a/\varepsilon_A - f_A(x_i,y_i)/\varepsilon_A))\\
~~~~~~~~~~~~.~\omega(b/\varepsilon_B - f_B(T_\mu((x_i,y_i))/\varepsilon_A))
\end{array}
\label{parzen}
\end{equation}

\noindent ‎where $\varepsilon_A$ and $\varepsilon_B$ are two positive scaling factors related to card ($L_A$) and card ($L_B$), respectively, in order to control the width of the Parzen window $\omega$. In this work, B-spline functions $\beta^n(x)$ are used as the Parzen window.
A B-spline of order $n$ can be generated by convolving the B-spline of order $0$ with itself $(n+1)$ times. This is a piece-wise polynomial of integer degree $n\geqslant0$ that can be recursively defined by the following convolution \cite{thevenaz2000}.
	\vspace{-0.1cm}
\begin{equation}‎
\begin{array}{l l}
\beta^n(x)=(\beta^{n-1}\ast \beta^0 )(x)\\
~~~~~~~=\displaystyle \int_{-\infty}^\infty \beta^{n-1}(x)\beta^0 (x-t)dt, ~~~n>0
\end{array}
\label{spline}
\end{equation}‎ 
	\vspace{-0.1cm}
\noindent where $\beta^0$ is a unit square pulse

\begin{equation}‎
\beta^0(x)=\Big\{ 
\begin{array}{l l}
1~~~;~~x\in[-0.5,0.5)\\
0~~~;~~x\notin [-0.5,0.5)
\end{array}.
\label{sign}
\end{equation}

The cubic B-spline function $\beta^3$ is selected as model kernel since it could provide a high-quality model.

Moreover, as $MI$ is sensitive to the amount of overlap between the images, Normalized Mutual Information NMI ($I^{'}$) in (\ref{NMI}) is used to overcome this problem \cite{pluim2000}. Here, any change in uncertainty of the image values and, therefore, the marginal entropies, will not result in a change in the alignment measure. Maximisation of $I^{'}$ seeks a transformation where joint entropy is minimised with respect to the marginal entropies.
	\vspace{-0.1cm}
\begin{equation}‎
I^{'}(\mu)=1+\frac {MI(\mu)}{E(a,b)}
\label{NMI}
\end{equation}

\noindent ‎where $E(a,b)$ is the joint entropy given in
	\vspace{-0.1cm}
\begin{equation}‎
E(a,b)=-\displaystyle \sum_{a\in L_A}\displaystyle \sum_{b\in L_B}p(a,b;\mu)log~p(a,b;\mu).
\label{jointEntropy}
\end{equation}‎

\subsubsection{Multiscale Framework}
A multiscale framework is a solution to solve the problem of high computation cost. The task is first solved at a coarse scale. Then, the results are propagated to the next finer level and used as a starting guess for solving the task at that level. This procedure is iterated until the finest level is reached. An image pyramid is made, a set of gradually reduced versions of the original image. Based on an image of size $M \times N$, a set of sequence of images are created with different sizes $\lbrace(M_1 \times N_1), ..., (M_i \times N_i), ..., (M \times N)\rbrace$ ‎where $M_i$ and $N_i$ are obtained from $M_i=\lfloor M_{i+1}/2 \rfloor$ and $N_i=\lfloor N_{i+1}/2 \rfloor$. It is very important that the initial condition for last level is the best possible in order to reduce the amount of refinement necessary to reach convergence. To get optimal starting conditions, it is crucial that the coarse levels of the pyramid most represent the finest level. For all cases, NMI between the whole overlap of subband images of two images is computed at each level and maximized successively, and the search is performed on an interval around the optimal transformation parameters found at the previous level and is refined at the next level. Thus, the accuracy of the search increases from coarse resolution to fine resolution. At each level, the iteration stops when the maximum number of 110 iterations is achieved, which are sufficient for the search to converge to a set of registration parameters with a subpixel accuracy based on our experiments.

\subsubsection{Search Strategy}
For the multiscale framework, a suitable optimizer is desired that converges in a few criterion evaluations when initialized with good starting conditions. The Levenberg-Marquardt (LM) algorithm \cite{marquardt1963} is an iterative technique that locates
the minimum of a multivariate function that is expressed as the sum of squares of non-linear real-valued functions. It has become a standard technique for non-linear least-squares problems, widely adopted in a broad spectrum of disciplines. LM can be thought of as a combination of steepest descent and the Gauss-Newton method. When the current solution is far from the correct one, the algorithm behaves like a steepest descent method: slow, but guaranteed to converge. When the current solution is close to the correct solution, it becomes a Gauss-Newton method.

The steepest-gradient descent is a minimization algorithm that can be succinctly described in (\ref{steepest}). Its local convergence is guaranteed, although it may be very slow.
\begin{equation}‎
\mu^{(k+1)}=\mu^{(k)}- \Gamma \nabla I^{'}(\mu^{(k)})
\label{steepest}
\end{equation}

\noindent where $\nabla I^{'}$ is the gradient of $I^{'}$ which is defined in (\ref{gradient}), a component of $\nabla I^{'}$ is also given in (\ref{component}).

\begin{equation}‎
\nabla I^{'}=\bigg[~\frac{\partial I^{'}}{\partial \mu_1}~,~\frac{\partial I^{'}}{\partial \mu_2}~,~...~\bigg]
\label{gradient}
\end{equation}‎

\begin{equation}‎
\frac{\partial I^{'}}{\partial \mu}=-\sum_{a\in L_A}\sum_{b\in L_B} \frac{\partial p(a,b;\mu)}{\partial \mu}~log_2(\frac{p(a,b;\mu)}{p_A(a;\mu)})
\label{component}
\end{equation}‎

A key problem is determination of the appropriate scaling diagonal matrix $\Gamma$. The Newton method is described by

\begin{equation}‎
\mu^{(k+1)}=\mu^{(k)}- (\nabla^2I^{'}(\mu^{(k)}))^{-1} \nabla I^{'}(\mu^{(k)})
\label{newton}
\end{equation}

\noindent ‎where $\nabla^2I^{'}$ is the Hessian of $I^{'}$, defined as the matrix of the second derivative of $I^{'}$ (\ref{derivative}); a component of $\nabla^2 I^{'}$ is also given in (\ref{component2}). Its convergence to an optimum is not guaranteed, it may converge to a saddle point. Even worse, it diverges from the desired solution when the problem is not convex.

\begin{equation}‎
\nabla^2I^{'} =
\begin{bmatrix}
\displaystyle\frac{\partial^2 I^{'}}{\partial \mu_1 \partial \mu_1} & \displaystyle\frac{\partial^2 I^{'}}{\partial \mu_1 \partial \mu_2} & \cdots \\
~&~&\\
\displaystyle\frac{\partial^2 I^{'}}{\partial \mu_2 \partial \mu_1} & \displaystyle\frac{\partial^2 I^{'}}{\partial \mu_2 \partial \mu_2} & \cdots  \\
~&~&\\
\vdots  & \vdots  & \ddots
\end{bmatrix}
\label{derivative}
\end{equation} 
\\
\begin{equation}‎
\begin{array}{l l}
\displaystyle\frac{\partial^2 I^{'}}{\partial \mu_1 \partial \mu_1}\approx  \frac{1}{log_e (2)} 
\bigg( \sum_{b\in L_B} \frac{\partial p_B(b)}{\partial \mu_1} \frac{\partial p_B(b)}{\partial \mu_2}\frac{1}{p_B(b)}\bigg)\\\\
~~- \displaystyle\frac{1}{log_e (2)}
\bigg( \sum_{a\in L_A}\sum_{b\in L_B} \frac{\partial p(a,b)}{\partial \mu_1} \frac{\partial p(a,b)}{\partial \mu_2}\frac{1}{p(a,b)}\bigg)
\end{array}
\label{component2}
\end{equation}‎

A modification of Marquardt–Levenberg proposed in \cite{thevenaz1997} exhibits superlinear convergence when close enough to the optimum. This specifically designed optimizer introduce a modified Hessian $HI^{'}$ (\ref{moddified}) in which we retain the off-diagonal entries of $\nabla^2I^{'}$ and multiply its diagonal entries by some factor. 
	\vspace{-0.1cm}
\begin{equation}‎
[HI^{'}(\mu)]_{i,j}=[\nabla^2I^{'}(\mu)]_{i,j}.(1+\delta_{i,j}\lambda)
\label{moddified} 
\end{equation}

\noindent ‎where $\delta_{i,j}$ is the Kronecker symbol, and where $\lambda$ is a tuning factor that represents the compromise between the gradient method and the Newton method. Therefore, the new update is
	\vspace{-0.1cm}
\begin{equation}‎
\mu^{(k+1)}=\mu^{(k)}- (HI^{'}(\mu^{(k)}))^{-1} \nabla I^{'}(\mu^{(k)})
\label{updated}
	\vspace{-0.5cm}
\end{equation}‎

In fact, search for a maximum of a function is nothing else than the search for the zeroes of the gradient of that function. Thus, above-mentioned steps are used to find the optimal parameters $\mu$ to maximise $I^{'}$ between an image and it's transformed image (obtained by the transform function $T_\mu$).
	\vspace{-0.2cm}
\subsection{Map Construction}
Once the transformation is applied to one of the images to take it to the co-ordinate system of the other
image, then the two images are aligned and can be stitched together to gradually construct the map. However, visible seams (due to exposure differences, misregistration, or moving objects) can occur during stitching process. In order to reduce such artefacts and achieve a smooth transition across the two images there should be a mechanism to decide which pixels to use and how to weight or blend them. In this work, in order to make the transition across images smooth, a simple but effective algorithm proposed in \cite{farbman2009} is used. In order to blend source image to the target image, let $c_1, c_2, ..., c_n$ be the $n$ points on the seam, $D_1, D_2, ..., D_n$ be the intensity differences at those points between the two images, and $b$ be a pixel of the source image. Then the intensity differences at pixel $b$ is computed as
	\vspace{-0.3cm}
\begin{equation}‎
D(b)=\sum_{i=1}^n w_i(b)D(c_i)
\label{blending1}
\end{equation}‎
\noindent where the weights $w_i(b)$ are
\begin{equation}‎
w_i(b)=\dfrac{1/\lVert c_i-b\rVert}{\sum_{j=1}^n 1/\lVert c_j-b\rVert}.
\label{blending2}
\end{equation}

\section{Results and Discussion} 
\label{resultsDiscussion}
In this section, results from experimental analysis are given to evaluate the overall performance of the proposed algorithm. Implementations have been carried out using MATLAB 8.3.0.532 on an Intel(R) Xeron(R) CPU E5-1620 0 @ 3.60GHz with 32GB of RAM. Two datasets are considered.

The first dataset concerns normal endoscopic images of the internal wall of the GI tract for two different subjects obtained by PillCam Capsule Endoscopy. The dataset of two different subjects totally contains almost 89800 colour images of size $256 \times 256$ in JPEG format.

\begin{figure}[!t]
	\centering
	\includegraphics[width=2.2in]{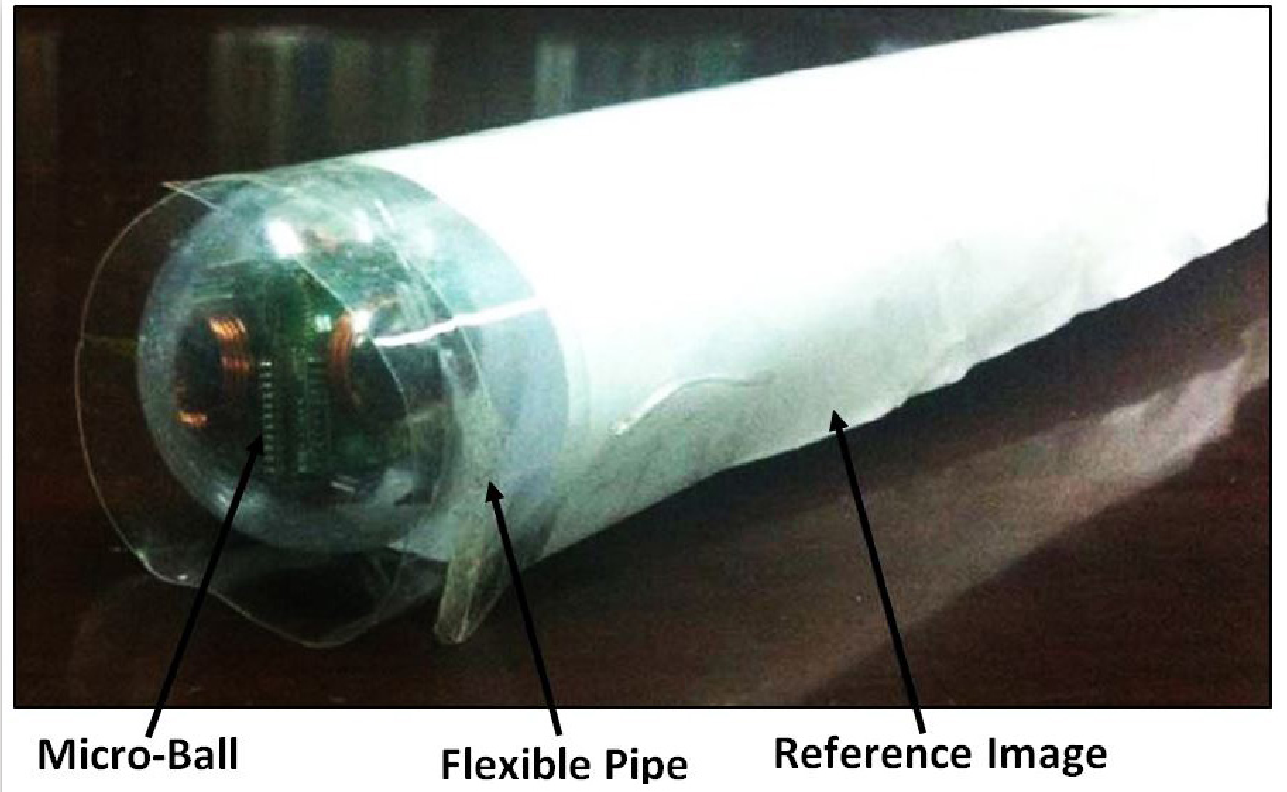}
	\caption{Experiment of moving the Micro-Ball through the cylinder.}
	\label{cylinder}
		\vspace{-0.5cm}
\end{figure}

The second dataset is based on the scenario which considers the simulation of the Micro-Ball performance, since the device is currently available as a prototype with a diameter of 4cm. The dataset in this scenario consists of 152 images of size $480 \times 480$ (in modified JPEG format with PSNR 40.7 dB and compression rate of 86\%) obtained from an experiment of moving the Micro-Ball through a cylinder. Although simulation of the surface deformation of GI tract can be neglected in this experiment (due to highly overlapped images and slow movement of the camera \cite{miranda2008}) surface deformation of GI tract and its impact on image stitching is considered in this work. The Micro-Ball is placed into a transparent, soft and flexible plastic pipe (see Fig. \ref{cylinder}) and images are captured while the Micro-Ball is moving and rotating through the slow-moving pipe. The outer wall of the pipe is covered with a reference picture. 
Similarly to the approach in \cite{froehlich2005}, observation of image sequences of the two subjects, are used to estimate the actual movement of the Micro-Ball inside the small bowel. The rotation occurrence frequency is approximately deduced from the perspective changes of the image sequences. The simulated movement is obtained by altering the translation frequency, average translation distance, and rotation occurrence frequency to adjust the prototype (40mm) with the actual size (15 mm). Translation frequency, average translation distance, and rotation occurrence frequency for real situation and simulation are 10 times/min, 3mm, 1 time/min and 10 times/min, 8mm, 1 time/min, respectively. 

In the following subsections, performance of the first two steps of the algorithm, namely keypoints detection and outlier removal  are examined, then, the accuracy of the proposed algorithm is verified quantitatively and visually to show how the algorithm would help to reach the goal of constructing a 2D reconstruction of the small intestinal tract. All results are based on the first dataset except Fig. \ref{microMap}.
	\vspace{-0.1cm}
\subsection{Comparison of scale invariant keypoints detectors} 
Outlier removal methods are set to eliminate false matches while preserving the correct ones. To select the optimum method, Correct Rate (CR) which is defined as follow is used, similar to \cite{mikolajczyk2005}. 
	\vspace{-0.2cm}
\begin{equation}‎
CR~(\%)=((c - f)/a)\times 100
\label{rate}
\end{equation}
where $c$, $f$, and $a$ are the number of remaining correct matches, remaining false matches, and all correct matches, respectively.

From the experimental results shown in Table~\ref{table_mismatch}, it follows:
1) MLESAC is more reliable than other methods (HT, RANSAC, and MSAC) for different keypoint detectors being capable of removing most of the false matches on the tested datasets. An illustrative example is shown in Fig. \ref{mismatches}, where the number of correct and false matches can be calculated statistically through visual inspection.
2) PCA-SIFT performs better than SIFT and SIFT+GC for the intended application. Thus, the combination of MLESAC and PCA-SIFT is a promising approach to detect the set of most reliable scale-invariant keypoints as the input that maximizes the NMI.
3) Observing the CR results in Table \ref{table_mismatch}, the correct rate has expectedly increased (3\% to 5\%) for all cases after applying the image enhancement method.

\begin{table}[!t]
	\renewcommand{\arraystretch}{1.3}
	\caption{Comparison of scale-invariant keypoints detectors using outlier removal algorithms, before and after enhancement.}
	\label{table_mismatch}
	\centering
	
	\begin{tabular}{ c|l|c|c }
		\multicolumn{2}{ c| }{Keypoints Detection}           & {\vtop{\hbox{\strut ~CR(\%)}\hbox{\strut{(before)}}}} & {\vtop{\hbox{\strut ~CR(\%)}\hbox{\strut{~(after)}}}} \\ \hline
		\multirow{4}{*}{SIFT}              & HT              &                        75.123                         &                        76.985                         \\
		                                   & RANSAC          &                        79.880                         &                        80.173                         \\
		                                   & MSAC            &                        84.671                         &                        85.971                         \\
		                                   & MLESAC          &                        88.013                         &                        88.995                         \\ \hline
		\multirow{4}{*}{\textbf{PCA-SIFT}} & HT              &                        79.801                         &                        81.764                         \\
		                                   & RANSAC          &                        83.875                         &                        88.452                         \\
		                                   & MSAC            &                        89.968                         &                        92.251                         \\
		                                   & \textbf{MLESAC} &                    \textbf{93.738}                    &                    \textbf{95.019}                    \\ \hline
		\multirow{4}{*}{SIFT+GC}           & HT              &                        76.123                         &                        78.311                         \\
		                                   & RANSAC          &                        79.747                         &                        81.090                         \\
		                                   & MSAC            &                        83.541                         &                        86.773                         \\
		                                   & MLESAC          &                        86.922                         &                        89.914                         \\ \hline
	\end{tabular}
		\vspace{-0.1cm}
\end{table}
	\vspace{-0.3cm}
\subsection{Image registration accuracy} 
To evaluate the registration performance, a ground truth is required to compare with the obtained registration parameters. Therefore, 2500 different pairs of images are generated by cropping and manipulating (rotating, scaling, shifting, shearing and projecting) the real endoscopic images (from the first dataset) with different overlap ratios. Thereafter, registration accuracy criterion is now computed as the Mean Euclidean Distance ($MED$) between homologous pixels of the target images and the registered images. The computation time (seconds) and registration accuracy in terms of the Mean and Standard Deviation (SD) for 2500 image pairs with different methods are shown in Table \ref{resulttotal}. 

\begin{table}[!t]
	\caption{Registration results obtained by running PCA-SIFT, NMI, and the Proposed method on 2500 different pairs of images.}
	\label{resulttotal}
	\centering
	\begin{tabular}{ c|c|c|c }
		Measure &  Method & Mean & SD  \\
		\hline
		\multirow{5}{*} {MED} 
		& PCA-SIFT		& 3.12  			& 0.98 \\
		& NMI 			& 2.68  			& 0.71 \\
		& Proposed		& \textbf{0.93}	& \textbf{0.33} \\ 		
		\hline
		\multirow{5}{*} {TIME} 
		& PCA-SIFT		& 6.87  			& 1.68 \\
		& NMI			& 10.54  			& 2.10 \\
		& Proposed		& \textbf{8.98}  & \textbf{0.98} \\ \hline
	\end{tabular}
			\vspace{-0.3cm}
\end{table}

It can be seen from the results in Table \ref{resulttotal} that the proposed algorithm shows a better performance compared with the results of either PCA-SIFT or NMI. Although the computation time has increased about two second on average, the accuracy has improved significantly compared to that of PCA-SIFT. Analogously, comparing with NMI, the proposed method shows a very good improvement in terms of time and accuracy.
	\vspace{-0.5cm}
\subsection{2D map construction} 
\begin{figure}[!t]
	\centering
	\includegraphics[width=3.2in]{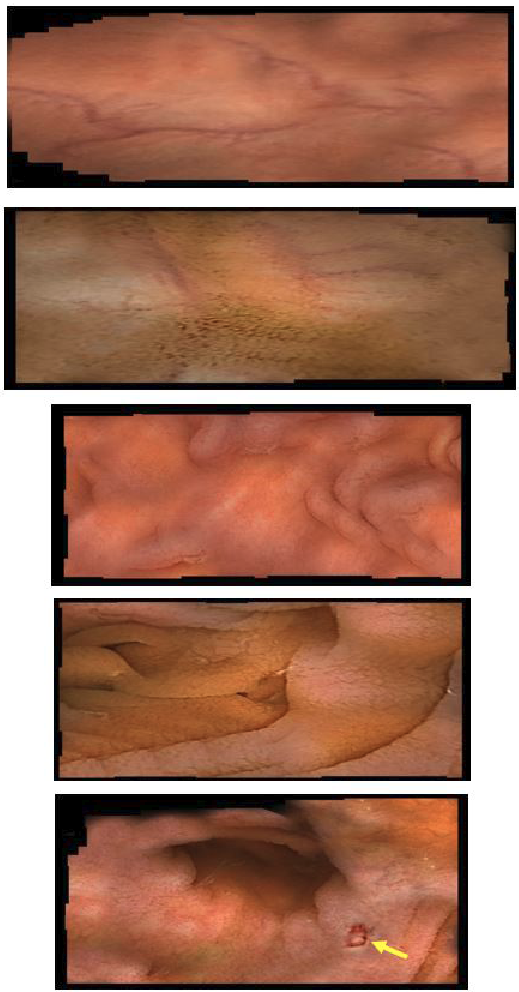}
	\caption{Five different regions of  GI tract are reconstructed by successful registration of 102, 78, 66, 86, and 98 successive images (top to bottom).}
	\label{wceMap}
		\vspace{-0.6cm}
\end{figure}

The proposed algorithm is run on sets of consecutive images obtained by the conventional wireless capsule endoscopy. In Fig. \ref{wceMap}, five segments of 2D representation of the stitched consecutive images are shown. Many cases are examined to generate a wider map; however, many regions are not covered by the cameras, only segments of the GI tract are able to  reconstruct a map, instead of one image of whole intestinal tract. This is because the wireless capsule endoscopy devices are equipped with 1 or 2 cameras, they can never provide enough information to enable reconstruction of a full view of the intestinal tract. The 2D reconstruction method relies on the overlapping areas of paired images and coverage rate of intestine's interior wall, the traditional wireless capsule devices could not provide enough information for the reconstruction of the whole intestinal tract due to their limited visual field and the negative effect of the random movement of the capsule. However, the Micro-Ball can achieve nearly 100\% coverage with its multiple cameras equipped with a smart imaging system. 

The abnormal region was found in Fig. \ref{wceMap} which reveals an arteriovenous malformation with active bleeding, confirmed by physicians, indicates the proposed method could perform well to show surroundings of an abnormal region for more efficient diagnosis.

\begin{figure}[!t]
	\centering
	\includegraphics[width=3in]{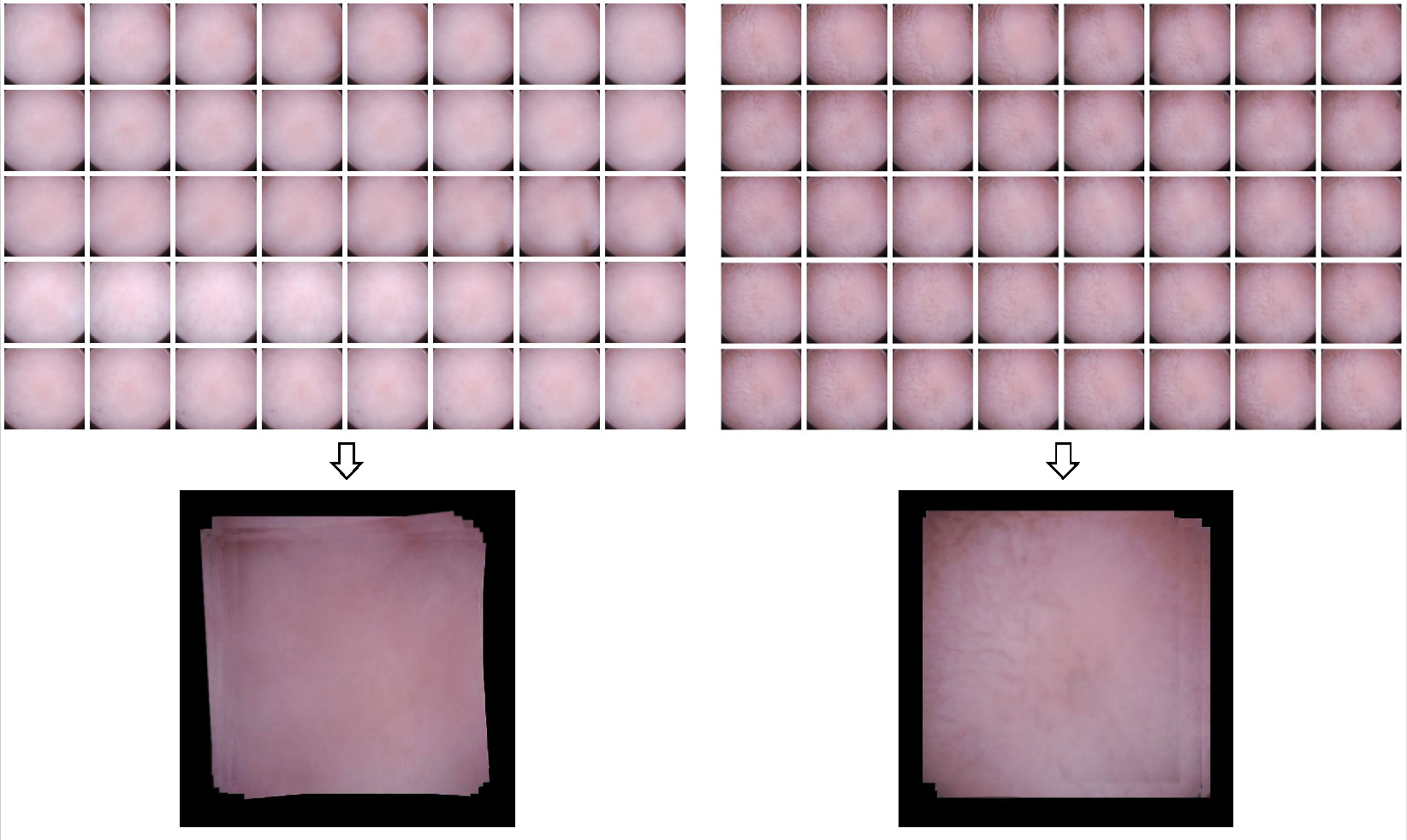}
	\caption{Two sample sets, each set with 40 consecutive images captured two different regions of intestinal wall. Images are registered by the proposed method, each image in second row represents the whole information of each image set in first row.}
	\label{redundants}
\end{figure}

The proposed algorithm is not able to create one continues map representing the whole area of GI tract as explained above, however, registration of any successive images that contain similar information is led to 64.89\% reduction in total number of images in the dataset.
Since there is no control on movement of wireless capsule endoscopy, its random movement causes producing a lot of redundant data when it stops moving, and still captures images from the same region repeatedly. The proposed algorithm by registering any similar adjacent images considerably reduces the number of images. For instance, Fig. \ref{redundants}, shows two sets of successive images of almost the same region of the small intestinal wall. Running the proposed algorithm on these two sets, the output will be one image that has all the information from the 40 images, as shown in Fig. \ref{redundants}.

\begin{figure}[!t]
	\centering
	\includegraphics[width=3in]{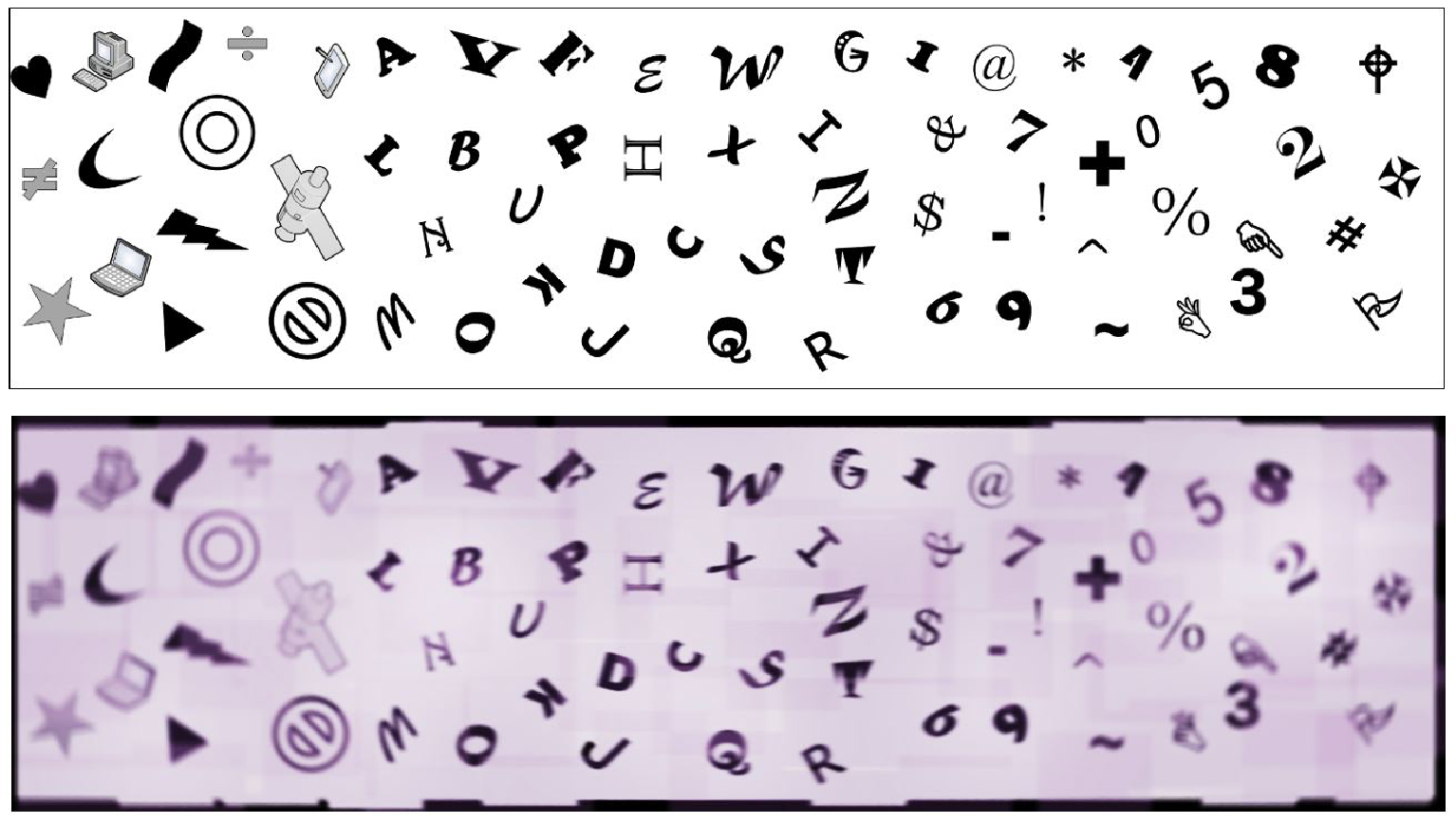}
	\caption{Reference picture which covers whole interior wall of the cylinder (top) and reconstructed map from 152 smaller images captured by Micro-Ball (bottom).}
	\label{microMap}
			\vspace{-0.4cm}
\end{figure}

Running the proposed method with the same implementation on the second dataset which is taken from moving the Micro-Ball through the cylinder, as shown in Fig. \ref{cylinder}, proves that the proposed method together with the Micro-Ball have been successful to make the entire field of view possible in 30.75 minutes with MED error accumulation of 16.59 pixels and without affecting the visual registration quality, compared with the reference image, as shown in Fig. \ref{microMap}. The Micro-Ball makes the impossible goal possible: One image represents whole information of the dataset. It provides a unique wide area visualization on which the observation and examination can be performed in an efficient way. Besides, from the clinical point of view, it is valuable to observe at once both the region of interest and its surrounding area, without passing from one image to another.

The importance of this work included but not limited to the following items:
First, this work proposes an efficient automatic algorithm that employs the robustness of feature-based methods and the accuracy of intensity-based methods to construct a hybrid scheme for endoscopic image registration. The results show that this arrangement of two different groups of image registration algorithm has been successful. Current version of the method works the best but someone may get better results in the future by changing the method in initialization step and/or fine-tuning step. The structure of the algorithm presented here is in a way that initialization step can be done with any feature-based method and then an intensity-based method is used to tune the transformation function. Second, the proposed registration strategy, besides common wireless endoscopic images, was tested on a unique database which has been obtained from \textit{in vitro} experiments with the Micro-Ball demo system. The algorithm has been successful to construct the whole inner wall of the \textit{in vitro} experiment and also many regions of the GI tract using real endoscopic images. The construction of the whole tract was not possible with real endoscopic images not because of the algorithm but due to limitations of the endoscopy devices i.e. high data missing rate, no location information, no smart imaging strategy, etc. These results are of significance for the challenging task of endoscopic image registration for different applications. Third, it should be noted that the aim of this work is not only to produce some improvement in the registration algorithm but the whole workflow is of high importance in the wireless endoscopy field. In fact, using the capability of the new developed smart Micro-Ball to construct the whole inner wall is also a promising result for the future of wireless endoscopy devices with multiple cameras.

\section{Conclusion and Future Work}
\label{conclusion}
A new fully automatic image registration approach has been proposed in this paper to register endoscopic images and construct a 2D representation of the small intestine's interior wall. The proposed method is based on maximization of normalised mutual information between any two consecutive images. Aiming at this, the initial solution for search strategy is obtained by the PCA-SIFT algorithm which is equipped with MLESAC outlier removal. Then, in a multiscale framework, the algorithm tries to find the best transformation parameters for registration and gradually form the whole map. The method has been examined on two different datasets and the registration accuracy and computation time are very satisfying.

For the future work, we plan to further reduce the computation time, by searching the most salient initial solution at the first stage without outlier removal algorithms and by adding high performance computing hardware such as GPU. Moreover, as the GI tract wall is a soft live object and its texture changes slightly during the examination, it is also important to consider the image deformation issue in endoscopic image registration. Another objective is to extend the methodology to process not every two images at a time but several images which may lead to higher quality reconstructed map.
\vspace{-0.4cm}

\section*{Acknowledgment}
This work was supported by: EU FP7 projects EYE2E (Grant No. 269118) and LIVCODE (Grant No. 295151); EU Horizon 2020 projects ENRICHME (643691) and STEP2DYNA (691154).

\section*{}
\bibliographystyle{ieeetr}
\bibliography{mybibfile}




\end{document}